\documentclass[a4paper,british]{IEEEtran}
\usepackage[T1]{fontenc}
\usepackage[latin9]{inputenc}
\usepackage{calc}
\usepackage{units}
\usepackage{amsmath}
\usepackage{graphicx}

\makeatletter

\pdfpageheight\paperheight
\pdfpagewidth\paperwidth

\usepackage{amsmath}
\usepackage{amssymb}

\usepackage{etoolbox}
\newcommand{\breedte}{p{0.4cm}}
\newcommand{\FullWidth}{p{2.8cm}}

\usepackage{graphicx}

\usepackage{lscape}

\newcommand{\Z}{Z_0}
\newcommand{\Y}{Y_0}
\newcommand{\T}{^\mathrm{N}}

\makeatother

\usepackage{babel}
\begin{document}

\title{Conversions between \\Electrical Network Representations}

\author{Adam Cooman (ELEC, VUB)}
\maketitle
\begin{abstract}
The behaviour of electrical networks can be described with many different
representations, each with their distinct benefits. In this paper,
we consider Z, Y, G, H, ABCD, S and T parameters. Formulas exist to
go from one representation to another, but implementing them is an
error-prone procedure. In this paper, we present a more elegant way
to implement the transformations based on matrix calculations.
\end{abstract}

Let us start by defining the different circuit parameters we will
consider in this paper.

\paragraph*{Impedance and Admittance parameters}

The behaviour of a circuit with $N$ ports is described by the following
relations

\vspace{-0.2cm}

\begin{equation}
\left[\begin{array}{c}
V_{1}\\
\vdots\\
V_{N}
\end{array}\right]=\mathbf{Z}\left[\begin{array}{c}
I_{1}\\
\vdots\\
I_{N}
\end{array}\right]\qquad\left[\begin{array}{c}
I_{1}\\
\vdots\\
I_{N}
\end{array}\right]=\mathbf{Y}\left[\begin{array}{c}
V_{1}\\
\vdots\\
V_{N}
\end{array}\right]\label{eq:ZYparams}
\end{equation}

where the $\mathbf{Y}$ and $\mathbf{Z}$ matrices are $N\times N$
matrices that contain the impedance parameters or admittance parameters
of the circuit.

\paragraph*{Mixed parameters}

When two-port circuits are considered, some specialised representations
exist that mix voltages and currents at different ports of the circuit. 

\vspace{-0.2cm}

\begin{equation}
\left[\begin{array}{c}
I_{1}\\
V_{2}
\end{array}\right]=\mathbf{G}\left[\begin{array}{c}
V_{1}\\
I_{2}
\end{array}\right]\qquad\left[\begin{array}{c}
V_{1}\\
I_{2}
\end{array}\right]=\mathbf{H}\left[\begin{array}{c}
I_{1}\\
V_{2}
\end{array}\right]\label{eq:GHparams}
\end{equation}

When the voltage and current at one of the ports is considered as
input signals, the ABCD parameters are obtained. Two different sets
of ABCD parameters are used here:

\vspace{-0.2cm}

\begin{equation}
\left[\begin{array}{c}
V_{1}\\
I_{1}
\end{array}\right]=\mathbf{A}\left[\begin{array}{c}
V_{2}\\
-I_{2}
\end{array}\right]\qquad\left[\begin{array}{c}
V_{2}\\
-I_{2}
\end{array}\right]=\mathbf{B}\left[\begin{array}{c}
V_{1}\\
I_{1}
\end{array}\right]\label{eq:ABparams}
\end{equation}

The $\mathbf{B}$ matrix is the inverse of the $\mathbf{A}$-matrix
if it exists.

\paragraph*{Wave-based parameters}

Instead of working with voltages and currents, incident and reflected
waves are used as inputs and outputs of the circuit representation.
The incident and reflected waves at a port are related to the voltages
and currents measured at the same port in the following way:

\textbf{
\begin{align}
A_{i} & =k\left(V_{i}+\Z I_{i}\right)\quad\quad B_{i}=k\left(V_{i}-\Z I_{i}\right)\label{eq:WaveTransfo}
\end{align}
}

where $Z_{0}$ is the normalisation impedance. The parameter $k$
depends on the preferred definition of the waves. Two different options
are available in literature \cite{Kurokawa1965,marks1992}:

\[
k=\frac{1}{2\sqrt{\Re\left\{ \Z\right\} }}\qquad k=\alpha\frac{\sqrt{\Re\left\{ \Z\right\} }}{2\left|\Z\right|}
\]

where $\Re\left\{ \Z\right\} $ indicates the real part of $\Z$ and
$\alpha$ is a free parameter of modulus 1.

The S-parameters are now defined as:

\vspace{-0.2cm}

\begin{equation}
\left[\begin{array}{c}
B_{1}\\
\vdots\\
B_{N}
\end{array}\right]=\mathbf{S}\left[\begin{array}{c}
A_{1}\\
\vdots\\
A_{N}
\end{array}\right]\label{eq:Sparams}
\end{equation}

where $\mathbf{S}$ is an $N\times N$ matrix that contains the S-parameters
of the circuit. The second wave-based representation we consider are
the T-parameters:

\vspace{-0.2cm}

\begin{equation}
\left[\begin{array}{c}
A_{1}\\
B_{1}
\end{array}\right]=\mathbf{T}\left[\begin{array}{c}
B_{2}\\
A_{2}
\end{array}\right]\label{eq:Tparams}
\end{equation}

\begin{figure}
\noindent \begin{centering}
\includegraphics{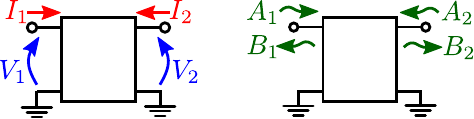}
\par\end{centering}

\caption{The currents are measured flowing into the circuit and all voltages
are referenced to a common ground node.}
\end{figure}

\section{Transforming between the representations}

All these different parameters describe the behaviour of the circuit.
Usually, the designer chooses the representation that best suits the
task at hand. It is therefore common to have to transform the circuit
representations from one into the other. There are some references
available that provide formulas to carry out these transformations,
but copying them is an error-prone task and many of the lists of conversion
formulas contain errors \cite{Frickey1994}. 

In general, the circuit parameters are in a certain representation
$\mathbf{R}$ which links the input signals $\mathbf{U}$ to output
signals $\mathbf{O}$:

\[
\mathbf{O}=\mathbf{R}\mathbf{U}
\]

They have to be transformed into another representation $\mathbf{R}\T$
with its inputs $\mathbf{U}\T$ and outputs $\mathbf{O}\T$

\begin{flalign*}
\mathbf{O}\T & =\mathbf{R}\T\mathbf{U}\T
\end{flalign*}

the transformation from one representation to another can be described
by looking at the transformation on the stacked input-output vectors:

\begin{equation}
\left[\begin{array}{c}
\mathbf{O}\T\\
\mathbf{U}\T
\end{array}\right]=\underbrace{\left[\begin{array}{cc}
\mathbf{P}_{11} & \mathbf{P}_{12}\\
\mathbf{P}_{21} & \mathbf{P}_{22}
\end{array}\right]}_{\mathbf{P}}\left[\begin{array}{c}
\mathbf{O}\\
\mathbf{U}
\end{array}\right]\label{eq:Generaltransfo}
\end{equation}

The goal of the transformation is to write $\mathbf{R}\T$ as a function
of the original representation $\mathbf{R}$ and the transformation
matrix $\mathbf{P}$. Solving (\ref{eq:Generaltransfo}) for $\mathbf{O}\T$
gives

\[
\mathbf{O}\T=\left(\mathbf{P}_{11}\mathbf{R}+\mathbf{P}_{12}\right)\left(\mathbf{P}_{21}\mathbf{R}+\mathbf{P}_{22}\right)^{-1}\mathbf{U}\T
\]

This give us the expression for $\mathbf{R}\T$ in function of $\mathbf{R}$:

\begin{equation}
\mathbf{R}\T=\left(\mathbf{P}_{11}\mathbf{R}+\mathbf{P}_{12}\right)\left(\mathbf{P}_{21}\mathbf{R}+\mathbf{P}_{22}\right)^{-1}\label{eq:TransfoFormula}
\end{equation}

This expression allows us to transform any representation into another
when we know the transformation matrix $\mathbf{P}$. The different
transformation matrices for most of the representations are listed
at the end of the paper. This approach is based on the excellent paper
on mixed-mode S-parameters where a similar transform is constructed
to transform from single-ended to mixed-mode \cite{Ferrero2006}.

\fbox{\begin{minipage}[t]{1\columnwidth}%
\vspace{0.2cm}

\subsection*{Example: Transforming from Z to G}

As an example, we show how we obtained the transformation matrix to
go from Z-parameters to G-parameters. For this specific transformation,
the following two stacked input-output vectors are obtained

\[
\left[\begin{array}{c}
\mathbf{O}\\
\mathbf{U}
\end{array}\right]=\left[\begin{array}{c}
V_{1}\\
V_{2}\\
I_{1}\\
I_{2}
\end{array}\right]\qquad\left[\begin{array}{c}
\mathbf{O}\T\\
\mathbf{U}\T
\end{array}\right]=\left[\begin{array}{c}
I_{1}\\
V_{2}\\
V_{1}\\
I_{2}
\end{array}\right]
\]

The transformation matrix $\mathbf{P}$ can now be found by finding
the permutation matrix that links the two vectors:

\[
\left[\begin{array}{c}
I_{1}\\
V_{2}\\
V_{1}\\
I_{2}
\end{array}\right]=\underbrace{\left[\begin{array}{cccc}
0_{\vphantom{1}} & 0 & 1 & 0\\
0_{\vphantom{1}} & 1 & 0 & 0\\
1_{\vphantom{1}} & 0 & 0 & 0\\
0_{\vphantom{1}} & 0 & 0 & 1
\end{array}\right]}_{\mathbf{P}}\left[\begin{array}{c}
V_{1}\\
V_{2}\\
I_{1}\\
I_{2}
\end{array}\right]
\]

Which results in the following conversion formula

\[
\mathbf{G}=\left(\left[\begin{array}{cc}
0 & 0\\
0 & 1
\end{array}\right]\mathbf{Z}+\left[\begin{array}{cc}
1 & 0\\
0 & 0
\end{array}\right]\right)\left(\left[\begin{array}{cc}
1 & 0\\
0 & 0
\end{array}\right]\mathbf{Z}+\left[\begin{array}{cc}
0 & 0\\
0 & 1
\end{array}\right]\right)^{-1}
\]

\vspace{0.2cm}
\end{minipage}}

\subsection*{Transforming into and from wave-based representations}

The transformation matrix $\mathbf{P}$ is not just a simple permutation
matrix when transforming from representations that use voltages and
currents to wave-based circuit representations and vice-versa. In
such transformations, voltages and currents are mixed to create the
waves, or incident and reflected waves are combined to obtain voltages
and currents. The formula to get the incident and reflected waves
from voltages and currents has been given before in (\ref{eq:WaveTransfo}).
Solving this equation for $V_{i}$ and $I_{i}$ gives the relations
between the voltage and current:

\[
\begin{cases}
\vphantom{\frac{1}{2k}}A_{i} & =k\left(V_{i}+\Z I_{i}\right)\\
\vphantom{\frac{1}{2k}}B_{i} & =k\left(V_{i}-\Z I_{i}\right)
\end{cases}\Leftrightarrow\begin{cases}
V_{i}= & \frac{1}{2k}\left(A_{i}+B_{i}\right)\\
I_{i}= & \frac{1}{2k}\left(\Y A_{i}-\Y B_{i}\right)
\end{cases}
\]

where $Y_{0}=\nicefrac{1}{Z_{0}}$. These formulas can now be used
to obtain the transformation matrix.

\fbox{\begin{minipage}[t]{1\columnwidth}%
\vspace{0.2cm}

\subsection*{Example: Transforming from S to Y}

Consider the transformation from S-parameters to Y-parameters. We
have the following stacked input-output vectors in this case

\[
\left[\begin{array}{c}
\mathbf{O}\\
\mathbf{U}
\end{array}\right]=\left[\begin{array}{c}
B_{1}\\
B_{2}\\
A_{1}\\
A_{2}
\end{array}\right]\qquad\left[\begin{array}{c}
\mathbf{O}\T\\
\mathbf{U}\T
\end{array}\right]=\left[\begin{array}{c}
I_{1}\\
I_{2}\\
V_{1}\\
V_{2}
\end{array}\right]
\]

so the transformation matrix $\mathbf{P}$ in this case is given by:

\[
\left[\begin{array}{c}
I_{1}\\
I_{2}\\
V_{1}\\
V_{2}
\end{array}\right]=\frac{1}{2k}\left[\begin{array}{cccc}
\Y & 0 & \Y & 0\\
0_{\vphantom{1}} & \Y & 0 & \Y\\
1_{\vphantom{1}} & 0 & 1 & 0\\
0_{\vphantom{1}} & 1 & 0 & 1
\end{array}\right]\left[\begin{array}{c}
B_{1}\\
B_{2}\\
A_{1}\\
A_{2}
\end{array}\right]
\]

which results in the following conversion formula:

\[
\mathbf{Y}=\left(\left[\begin{array}{cc}
\Y & 0\\
0 & \Y
\end{array}\right]\mathbf{S}+\left[\begin{array}{cc}
\Y & 0\\
0 & \Y
\end{array}\right]\right)\left(\left[\begin{array}{cc}
1 & 0\\
0 & 1
\end{array}\right]\mathbf{S}+\left[\begin{array}{cc}
1 & 0\\
0 & 1
\end{array}\right]\right)^{-1}
\]

\vspace{0.2cm}
\end{minipage}}

\section{Conclusion}

A very simple-to implement method is obtained to go from one circuit
representation to another. The method uses a transformation matrix
$\mathbf{P}$, which is listed for most of the transformations at
the end of the paper. The actual transform boils down to splitting
$\mathbf{P}$ in four parts

\[
\mathbf{P}=\left[\begin{array}{cc}
\mathbf{P}_{11} & \mathbf{P}_{12}\\
\mathbf{P}_{21} & \mathbf{P}_{22}
\end{array}\right]
\]

and then calculating

\[
\mathbf{R}\T=\left(\mathbf{P}_{11}\mathbf{R}+\mathbf{P}_{12}\right)\left(\mathbf{P}_{21}\mathbf{R}+\mathbf{P}_{22}\right)^{-1}
\]

where $\mathbf{R}$ is the matrix describing the circuit in its original
representation and $\mathbf{R}\T$ is the circuit matrix in the new
representation.

\vfill{}

\newpage{}

\begin{landscape}
%
%
%
%
%
%
%
%

\renewcommand{\breedte}{p{0.41cm}}
\renewcommand{\FullWidth}{p{2.8cm}}

\renewcommand{\Z}{\ensuremath{\! Z_{\! 0}}}
\renewcommand{\Y}{\ensuremath{\! Y_{\! 0}}}
\newcommand{\heightcontrol}{\rule[-1.1cm]{0mm}{23mm}}

\begin{tabular}{c||\FullWidth|\FullWidth|\FullWidth|\FullWidth|\FullWidth|\FullWidth|\FullWidth}%
$from\backslash to$ & \begin{center}$\mathbf{Y}$~\eqref{eq:ZYparams}\end{center} & \begin{center}$\mathbf{Z}$~\eqref{eq:ZYparams}\end{center}  & \begin{center}$\mathbf{G}$~\eqref{eq:GHparams}\end{center}  & \begin{center}$\mathbf{H}$~\eqref{eq:GHparams}\end{center}  & \begin{center}$\mathbf{A}$~\eqref{eq:ABparams}\end{center}  & \begin{center}$\mathbf{S}$~\eqref{eq:Sparams}\end{center}  & \begin{center}$\mathbf{T}$~\eqref{eq:Tparams}\end{center}  \\%
\hline %
\hline %
\heightcontrol$\mathbf{Y}$~\eqref{eq:ZYparams} &  & $\hphantom{\frac{1}{2k}}\left[\begin{array}{\breedte\breedte\breedte\breedte}%
0 & 0 & 1 & 0\\%
0 & 0 & 0 & 1\\%
1 & 0 & 0 & 0\\%
0 & 1 & 0 & 0%
\end{array}\right]$ & $\hphantom{\frac{1}{2k}}\left[\begin{array}{\breedte\breedte\breedte\breedte}%
1 & 0 & 0 & 0\\%
0 & 0 & 0 & 1\\%
0 & 0 & 1 & 0\\%
0 & 1 & 0 & 0%
\end{array}\right]$ & $\hphantom{\frac{1}{2k}}\left[\begin{array}{\breedte\breedte\breedte\breedte}%
0 & 0 & 1 & 0\\%
0 & 1 & 0 & 0\\%
1 & 0 & 0 & 0\\%
0 & 0 & 0 & 1%
\end{array}\right]$ & $\hphantom{\frac{1}{2k}}\left[\begin{array}{\breedte\breedte\breedte\breedte}%
0 & 0 & 1 & 0\\%
1 & 0 & 0 & 0\\%
0 & 0 & 0 & 1\\%
0 & -1 & 0 & 0%
\end{array}\right]$ & $k\left[\begin{array}{\breedte\breedte\breedte\breedte}%
-\Z & 0 & 1 & 0\\%
0 & -\Z & 0 & 1\\%
\Z & 0 & 1 & 0\\%
0 & \Z & 0 & 1%
\end{array}\right]$ & $k\left[\begin{array}{\breedte\breedte\breedte\breedte}%
0 & -\Z & 0 & 1\\%
0 & \Z & 0 & 1\\%
\Z & 0 & 1 & 0\\%
-\Z & 0 & 1 & 0%
\end{array}\right]$  \\
\hline

\heightcontrol$\mathbf{Z}$~\eqref{eq:ZYparams} & $\hphantom{\frac{1}{2k}}\left[\begin{array}{\breedte\breedte\breedte\breedte}%
0 & 0 & 1 & 0\\%
0 & 0 & 0 & 1\\%
1 & 0 & 0 & 0\\%
0 & 1 & 0 & 0%
\end{array}\right]$ &  & $\hphantom{\frac{1}{2k}}\left[\begin{array}{\breedte\breedte\breedte\breedte}%
0 & 0 & 1 & 0\\%
0 & 1 & 0 & 0\\%
1 & 0 & 0 & 0\\%
0 & 0 & 0 & 1%
\end{array}\right]$ & $\hphantom{\frac{1}{2k}}\left[\begin{array}{\breedte\breedte\breedte\breedte}%
1 & 0 & 0 & 0\\%
0 & 0 & 0 & 1\\%
0 & 0 & 1 & 0\\%
0 & 1 & 0 & 0%
\end{array}\right]$ & $\hphantom{\frac{1}{2k}}\left[\begin{array}{\breedte\breedte\breedte\breedte}%
1 & 0 & 0 & 0\\%
0 & 0 & 1 & 0\\%
0 & 1 & 0 & 0\\%
0 & 0 & 0 & -1%
\end{array}\right]$ & $k\left[\begin{array}{\breedte\breedte\breedte\breedte}%
1 & 0 & -\Z & 0\\%
0 & 1 & 0 & -\Z\\%
1 & 0 & \Z & 0\\%
0 & 1 & 0 & \Z
\end{array}\right]$ & $k\left[\begin{array}{\breedte\breedte\breedte\breedte}%
0 & 1 & 0 & -\Z\\%
0 & 1 & 0 & \Z\\%
1 & 0 & \Z & 0\\%
1 & 0 & -\Z & 0%
\end{array}\right]$ \\%
\hline%
\heightcontrol$\mathbf{G}$~\eqref{eq:GHparams} & $\hphantom{\frac{1}{2k}}\left[\begin{array}{\breedte\breedte\breedte\breedte}%
1 & 0 & 0 & 0\\%
0 & 0 & 0 & 1\\%
0 & 0 & 1 & 0\\%
0 & 1 & 0 & 0%
\end{array}\right]$ & $\hphantom{\frac{1}{2k}}\left[\begin{array}{\breedte\breedte\breedte\breedte}%
0 & 0 & 1 & 0\\%
0 & 1 & 0 & 0\\%
1 & 0 & 0 & 0\\%
0 & 0 & 0 & 1%
\end{array}\right]$ &  & $\hphantom{\frac{1}{2k}}\left[\begin{array}{\breedte\breedte\breedte\breedte}%
0 & 0 & 1 & 0\\%
0 & 0 & 0 & 1\\%
1 & 0 & 0 & 0\\%
0 & 1 & 0 & 0%
\end{array}\right]$ & $\hphantom{\frac{1}{2k}}\left[\begin{array}{\breedte\breedte\breedte\breedte}%
0 & 0 & 1 & 0\\%
1 & 0 & 0 & 0\\%
0 & 1 & 0 & 0\\%
0 & 0 & 0 & -1%
\end{array}\right]$ & $k\left[\begin{array}{\breedte\breedte\breedte\breedte}%
-\Z & 0 & 1 & 0\\%
0 & 1 & 0 & -\Z\\%
\Z & 0 & 1 & 0\\%
0 & 1 & 0 & \Z%
\end{array}\right]$ & $k\left[\begin{array}{\breedte\breedte\breedte\breedte}%
0 & 1 & 0 & -\Z\\%
0 & 1 & 0 & \Z\\%
\Z & 0 & 1 & 0\\%
-\Z & 0 & 1 & 0%
\end{array}\right]$ \\%
\hline%
\heightcontrol$\mathbf{H}$~\eqref{eq:GHparams} & $\hphantom{\frac{1}{2k}}\left[\begin{array}{\breedte\breedte\breedte\breedte}%
0 & 0 & 1 & 0\\%
0 & 1 & 0 & 0\\%
1 & 0 & 0 & 0\\%
0 & 0 & 0 & 1%
\end{array}\right]$ & $\hphantom{\frac{1}{2k}}\left[\begin{array}{\breedte\breedte\breedte\breedte}%
1 & 0 & 0 & 0\\%
0 & 0 & 0 & 1\\%
0 & 0 & 1 & 0\\%
0 & 1 & 0 & 0%
\end{array}\right]$ & $\hphantom{\frac{1}{2k}}\left[\begin{array}{\breedte\breedte\breedte\breedte}%
0 & 0 & 1 & 0\\%
0 & 0 & 0 & 1\\%
1 & 0 & 0 & 0\\%
0 & 1 & 0 & 0%
\end{array}\right]$ &  & $\hphantom{\frac{1}{2k}}\left[\begin{array}{\breedte\breedte\breedte\breedte}%
1 & 0 & 0 & 0\\%
0 & 0 & 1 & 0\\%
0 & 0 & 0 & 1\\%
0 & -1 & 0 & 0%
\end{array}\right]$ & $k\left[\begin{array}{\breedte\breedte\breedte\breedte}%
1 & 0 & -\Z & 0\\%
0 & -\Z & 0 & 1\\%
1 & 0 & \Z & 0\\%
0 & \Z & 0 & 1%
\end{array}\right]$ & $k\left[\begin{array}{\breedte\breedte\breedte\breedte}%
0 & -\Z & 0 & 1\\%
0 & \Z & 0 & 1\\%
1 & 0 & \Z & 0\\%
1 & 0 & -\Z & 0%
\end{array}\right]$ \\%
\hline%
\heightcontrol$\mathbf{A}$~\eqref{eq:ABparams} & $\hphantom{\frac{1}{2k}}\left[\begin{array}{\breedte\breedte\breedte\breedte}%
0 & 1 & 0 & 0\\%
0 & 0 & 0 & -1\\%
1 & 0 & 0 & 0\\%
0 & 0 & 1 & 0%
\end{array}\right]$ & $\hphantom{\frac{1}{2k}}\left[\begin{array}{\breedte\breedte\breedte\breedte}%
1 & 0 & 0 & 0\\%
0 & 0 & 1 & 0\\%
0 & 1 & 0 & 0\\%
0 & 0 & 0 & -1%
\end{array}\right]$ & $\hphantom{\frac{1}{2k}}\left[\begin{array}{\breedte\breedte\breedte\breedte}%
0 & 1 & 0 & 0\\%
0 & 0 & 1 & 0\\%
1 & 0 & 0 & 0\\%
0 & 0 & 0 & -1%
\end{array}\right]$ & $\hphantom{\frac{1}{2k}}\left[\begin{array}{\breedte\breedte\breedte\breedte}%
1 & 0 & 0 & 0\\%
0 & 0 & 0 & -1\\%
0 & 1 & 0 & 0\\%
0 & 0 & 1 & 0%
\end{array}\right]$ &  & $k\left[\begin{array}{\breedte\breedte\breedte\breedte}%
1 & -\Z & 0 & 0\\%
0 & 0 & 1 & \Z\\%
1 & \Z & 0 & 0\\%
0 & 0 & 1 & -\Z%
\end{array}\right]$ & $k\left[\begin{array}{\breedte\breedte\breedte\breedte}%
0 & 0 & 1 & \Z\\%
0 & 0 & 1 & -\Z\\%
1 & \Z & 0 & 0\\%
1 & -\Z & 0 & 0%
\end{array}\right]$ \\%
\hline%
\heightcontrol$\mathbf{S}$~\eqref{eq:Sparams} & $\frac{1}{2k}\left[\begin{array}{\breedte\breedte\breedte\breedte}%
-\Y & 0 & \Y & 0\\%
0 & -\Y & 0 & \Y\\%
1 & 0 & 1 & 0\\%
0 & 1 & 0 & 1%
\end{array}\right]$ & $\frac{1}{2k}\left[\begin{array}{\breedte\breedte\breedte\breedte}%
1 & 0 & 1 & 0\\%
0 & 1 & 0 & 1\\%
-\Y & 0 & \Y & 0\\%
0 & -\Y & 0 & \Y%
\end{array}\right]$ & $\frac{1}{2k}\left[\begin{array}{\breedte\breedte\breedte\breedte}%
-\Y & 0 & \Y & 0\\%
0 & 1 & 0 & 1\\%
1 & 0 & 1 & 0\\%
0 & -\Y & 0 & \Y%
\end{array}\right]$ & $\hphantom{\frac{1}{2k}}\left[\begin{array}{\breedte\breedte\breedte\breedte}%
1 & 0 & 1 & 0\\%
0 & -\Y & 0 & \Y\\%
-\Y & 0 & \Y & 0\\%
0 & 1 & 0 & 1%
\end{array}\right]$ & $\frac{1}{2k}\left[\begin{array}{\breedte\breedte\breedte\breedte}%
1 & 0 & 1 & 0\\%
-\Y & 0 & \Y & 0\\%
0 & 1 & 0 & 1\\%
0 & \Y & 0 & -\Y%
\end{array}\right]$ &  & $\hphantom{k}\left[\begin{array}{\breedte\breedte\breedte\breedte}%
0 & 1 & 0 & 0\\%
0 & 0 & 0 & 1\\%
0 & 0 & 1 & 0\\%
1 & 0 & 0 & 0%
\end{array}\right]$ \\%
\hline%
\heightcontrol$\mathbf{T}$~\eqref{eq:Tparams} & $\frac{1}{2k}\left[\begin{array}{\breedte\breedte\breedte\breedte}%
0 & 0 & \Y & -\Y\\%
-\Y & \Y & 0 & 0\\%
0 & 0 & 1 & 1\\%
1 & 1 & 0 & 0%
\end{array}\right]$ & $\frac{1}{2k}\left[\begin{array}{\breedte\breedte\breedte\breedte}%
0 & 0 & 1 & 1\\%
1 & 1 & 0 & 0\\%
0 & 0 & \Y & -\Y\\%
-\Y & \Y & 0 & 0%
\end{array}\right]$ & $\frac{1}{2k}\left[\begin{array}{\breedte\breedte\breedte\breedte}%
0 & 0 & \Y & -\Y\\%
1 & 1 & 0 & 0\\%
0 & 0 & 1 & 1\\%
-\Y & \Y & 0 & 0%
\end{array}\right]$ & $\frac{1}{2k}\left[\begin{array}{\breedte\breedte\breedte\breedte}%
0 & 0 & 1 & 1\\%
-\Y & \Y & 0 & 0\\%
0 & 0 & \Y & -\Y\\%
1 & 1 & 0 & 0%
\end{array}\right]$ & $\frac{1}{2k}\left[\begin{array}{\breedte\breedte\breedte\breedte}%
0 & 0 & 1 & 1\\%
0 & 0 & \Y & -\Y\\%
1 & 1 & 0 & 0\\%
\Y & -\Y & 0 & 0%
\end{array}\right]$ & $\hphantom{k}\left[\begin{array}{\breedte\breedte\breedte\breedte}%
0 & 0 & 0 & 1\\%
1 & 0 & 0 & 0\\%
0 & 0 & 1 & 0\\%
0 & 1 & 0 & 0%
\end{array}\right]$ & \\%
\end{tabular}

\end{landscape}

\end{document}